\documentclass[twocolumn]{aastex61}

\usepackage{hyperref} 
\usepackage{amsmath}

\usepackage{mathrsfs}

\usepackage[]{units} 

\usepackage{threeparttable}
\usepackage{multirow}

\usepackage{soul}

\usepackage{ulem}

\usepackage{graphicx}				
\usepackage{dblfloatfix}

\begin{document}

\title{Following the Cosmic Evolution of Pristine Gas II: The search for Pop III--Bright Galaxies }
\shorttitle{The search for Pop III--bright Galaxies}
\author{Richard Sarmento}
\author{Evan Scannapieco}
\author{Seth Cohen}
\affil{School of Earth and Space Exploration, 
Arizona State University, 
P.O. Box 871404, Tempe, AZ, 
85287-1404
}

\shortauthors{Sarmento et al.}

\begin{abstract}
Direct observational searches for Population III (Pop III) stars at high redshift are faced with the question of how to select the most promising targets for spectroscopic follow-up. To help answer this, we use a large-scale cosmological simulation, augmented with a new subgrid model that tracks the fraction of pristine gas, to follow the evolution of high-redshift galaxies and the Pop III stars they contain.  We generate rest-frame ultraviolet (UV) luminosity functions for our galaxies and find that they are consistent with current $z \ge 7 $ observations. Throughout the redshift range $7 \le z \le 15$  we identify ``Pop III--bright'' galaxies  as those with at least 75\% of their flux coming from Pop III stars. 
While less than 1\% of galaxies brighter than $m_{\rm UV, AB} = 31.4$  mag are Pop III--bright in the range $7\leq z \leq8$, roughly 17\% of such galaxies are Pop III--bright at $z=9$, immediately before reionization occurs in our simulation. Moving to $z=10$, $m_{\rm UV, AB} = 31.4$  mag corresponds to larger, more luminous galaxies and the Pop III--bright fraction falls off to 5\%. Finally, at the highest redshifts, a large fraction (29\% at $z=14$ and 41\% at $z=15)$ of all galaxies are Pop III--bright regardless of magnitude. While $m_{\rm UV, AB} = 31.4$  mag galaxies are extremely rare during this epoch, we find that 13\% of galaxies at $z = 14$ are Pop III--bright with $m_{\rm UV, AB} \le 33$ mag, an intrisic magnitude within reach of the \textit{James Webb Space Telescope} using lensing. Thus, we predict that the best redshift to search for luminous Pop III--bright galaxies is just before reionization, while lensing surveys for fainter galaxies should push to the highest redshifts possible.
\end{abstract}



\section{Introduction}
Finding and characterizing the first galaxies is the next frontier in observational astronomy. Theoretical studies suggest that these metal-free stars could be observed today if their initial mass function (IMF) extended to low masses \citep{2006ApJ...653..285S, 2006ApJ...641....1T, 2007ApJ...661...10B,2010MNRAS.401L...5S,2015MNRAS.447.3892H,2016arXiv160200465I}. However, no one has yet observed a Population III (Pop III) star in or near the Galaxy \citep{2002Natur.419..904C, 2004A&A...416.1117C, 2006ApJ...639..897A,2005Natur.434..871F, 2007ApJ...670..774N,2011Natur.477...67C,2014Natur.506..463K,2015Natur.527..484H}. 

High-redshift observations have yielded candidates for Pop III stellar populations \citep{2002ApJ...565L..71M, 2004ApJ...617..707D, 2006Natur.440..501J, 2007MNRAS.379.1589D, 2008ApJ...680..100N, 2012ApJ...761...85K, 2013A&A...556A..68C}, without definitive detections. 
These include a controversial z = 6.6 galaxy analyzed by \cite{2015ApJ...808..139S} that displays He II $\lambda$1640 emission -- an indicator of the hard-ultraviolet (UV) spectrum produced by Pop III stars \citep{2001ApJ...550L...1T}. Yet, to date, there has not been a confirmed observation of a galaxy dominated by the flux from Pop III stars \citep{2016arXiv160900727B,2017MNRAS.468L..77P}.  

This may change in the near future. The soon-to-launch \textit{James Webb Space Telescope} (JWST) is poised to greatly expand our understanding of the high-redshift universe and possibly detect the first galaxies dominated by Pop III flux. Using the JWST, astronomers will be able to assemble galaxy catalogs out to $z=10$ and beyond and probe the era of the first galaxies \citep{2016jwst}.  However, planning for such observations requires estimating how such galaxies are distributed and, even more importantly, what fraction of galaxies as a function of magnitude and redshift will be dominated by Pop III flux -- warranting spectroscopic follow-up.

For now we only have general observational clues about the history of such early galaxy formation.
Using extremely deep Hubble Space Telescope (HST) observations, astronomers have been able to amass photometric galaxy catalogs out to $z=8$ and place initial constraints on galaxy populations out to $z\approx11$ \citep{2017arXiv170204867I, 2016PASA...33...37F, 2015ApJ...803...34B, 2016ApJ...816...46M, 2015MNRAS.450.3032M, 2013ApJ...762...32C, 2013ApJ...773...75O}. While a lot of progress has been made, the latest work at $z>8$ is hampered by small number statistics and completeness uncertainties \citep{2017ApJ...835..113L, 2015ApJ...808..104O, 2015ApJ...814...69A}.  

Several groups have used large-scale cosmological simulations and analytic models to investigate galaxy formation, the high-$z$ luminosity function (LF), and galaxy assembly \citep{2012MNRAS.423.1992S, 2015ApJ...807L..12O, 2016ApJ...816...46M,2017arXiv170102749B}. Others have used simulations to explore the transition between Pop III and Population II (Pop II) star formation \citep{2003ApJ...589...35S, 2007MNRAS.382..945T, 2007ApJ...654...66O, 2010ApJ...712..435T, 2010MNRAS.407.1003M, 2011ApJ...740...13Z, 2012ApJ...745...50W, 2013ApJ...773..108C, 2013MNRAS.428.1857J, 2013ApJ...775..111P,2014MNRAS.440.2498P}.

By definition, the first generation of Pop III stars must have formed in the primordial gas. However, an IMF lacking low-mass stars may also result from gas with metallicity below a critical threshold, $Z_{\rm crit}$. The exact value of the threshold depends on whether the dominant cooling channel for the gas is the fine-structure lines of metals or dust emission  \citep{2003Natur.422..869S, 2003Natur.425..812B,2005ApJ...626..627O}. While the value is poorly constrained, it is believed to be in the range  $10^{-6} < Z_{\rm crit} < 10^{-3} Z_{\odot}$. 
 
Here we make use of the work described in \cite{2017ApJ...834...23S} to track the pollution of the pristine gas at subgrid scales in high-resolution simulations of galaxy formation at high redshift. By following the evolution of the pristine gas, we can estimate the fraction of Pop III stars created in regions that would otherwise be considered polluted above $Z_{\rm crit}$.  
This allows us to present theoretical predictions for deep photometric galaxy surveys and, in particular, to characterize the fraction of Pop III flux in early galaxies. 
This information can guide planning for spectrographic follow-up in the search for Pop III stars, searching for their unique observational characteristics  \citep{2015MNRAS.450.2506V}. 


Our approach uses a customized version of \textsc{ramses} \citep{2002A&A...385..337T}, a cosmological adaptive mesh refinement (AMR) code, to follow galaxy formation from the dawn of star formation, at $z\approx 21$, to $z=7$.   
Using these simulation results, we generate rest-frame UV ($1500\AA$) galaxy luminosity functions, to demonstrate that our approach is consistent with existing photometric surveys and generate higher-redshift galaxy LFs for a set of JWST Near InfraRed Camera (NIRCam) filters to aid in planning for future such surveys.

Furthermore, using our unique capability to track the rate of subgrid metal pollution, we trace the formation of Pop III stars in these early galaxies and model their impact on the galaxies' flux. In doing so, we are able to identify a fraction of galaxies across a range of redshifts that have a significant fraction of Pop III stellar  flux.  This allows us to make predictions as to the galaxy luminosities and redshifts that are most likely to show Pop III features, such as narrow He II $\lambda$1640 emission, when they are followed up spectroscopically.

The work is structured as follows.  In Section 2 we describe our methods, including a brief discussion of the implementation of our subgrid model for following the evolution of the pristine gas fraction, our approach to halo finding, and the spectral energy distribution (SED) models used to compute the luminosity of our stars. In Section 3 we show that our high-redshift LF agrees with current observations and make predictions for future JWST surveys. Next, we focus on an analysis of the fraction of Pop III flux emitted by early galaxies that can be used to guide the search for metal-free stars. Conclusions are discussed in Section 4. 

\section{Methods}

\subsection{Simulation Setup and Characteristics}

We adopt the following cosmological parameters $\Omega_{\rm M} = 0.267$, $\Omega_{\Lambda} = 0.733$, $\Omega_{\rm b} = 0.0449$, $h = 0.71$, $\sigma_8 = 0.801$, and $n = 0.96,$ based on \cite{2011ApJS..192...18K}, where $\Omega_{\rm M}$, $\Omega_{\Lambda}$, and $\Omega_{\rm b}$ are the total matter, vacuum, and baryonic densities, respectively, in units of the critical density; $h$ is the Hubble constant in units of 100 km/s; $\sigma_8$ is the variance of linear fluctuations on the 8 $h^{-1}$ Mpc scale; and $n$ is the ``tilt" of the primordial power spectrum \citep{2011ApJS..192...16L}. 

For this study, we make use of \textsc{ramses} \citep{2002A&A...385..337T}, a cosmological adaptive mesh refinement (AMR) simulation code that uses an unsplit second-order Godunov scheme for evolving the Euler equations. \textsc{Ramses} tracks cell-centered variables that are interpolated to the cell faces for flux calculations. Flux between cells is computed using a Harten--Lax--van Leer--Contact Riemann solver \citep{1979JCoPh..32..101V,1988SJNA...25..294E} and the code is capable of advecting any number of these scalar quantities across simulation cells.  Self-gravity is solved using the multigrid method \citep{GT2011} along with the conjugate gradient method for levels $\ge 12$ in our simulation.  Stars and DM are modeled with collisionless particles and are evolved using a particle-mesh solver with cloud-in-cell interpolation.

We use \textsc{ramses} to evolve a 12 Mpc $h^{-1}$ on-a-side volume from Multi-Scale Initial Conditions (MUSIC) \citep{2013ascl.soft11011H} generated initial conditions through $z=7$. The initial gas metallicity was $Z = 0,$ the initial $H_{\rm 2}$ fraction was $10^{-6}$ \citep{2005MNRAS.363..393R}, and we define $Z_{\rm crit} = 10^{-5} Z_\odot.$  The base resolution of $1024^{3}$ cells ($\textit{l}_{\rm min} = 10$) corresponds to a grid resolution of 11.7 comoving kpc h$^{-1}$, and a dark matter (DM) particle mass of $4.47 \times 10^{5}\, M_{\odot}\, h^{-1}\, \Omega_{\rm \textsc{dm}}.$ We refine cells as they become $8\times$ overdense, resulting in a quasi-Lagrangian approach to refinement. We allowed for up to eight additional refinement levels ($\textit{l}_{\rm max} = 18$), resulting in an average physical spatial resolution of 45.8 pc h$^{-1}$.  Our choice of parameters resulted in a range of star particle masses $8.6\times10^3\, M_{\odot} \leq M_{\star} \leq 6.2\times 10^4\, M_{\odot}$.  The highest refinement level reached was 15. The nonlinear length scale at the end of the simulation, $z= 7$, was $47$ comoving kpc h$^{-1}$, corresponding to a mass of $3.2\times 10^{7} M_\odot$ h$^{-1}$. 
We did not model sink particles (black holes (BH)) in our simulation since BH feedback is not likely to be significant for our very early galaxies \citep{2008MNRAS.391..481S, 2004ApJ...608...62S}.  We tune the code reionization parameters to ensure that the reionization redshift occurs at $z_{\rm reion} \approx 8.5$, as reported by the \cite{2015arXiv150201589P}. Finally, all magnitudes are in the AB system \citep{1983ApJ...266..713O}.

\subsection{Simulation Physics}

Cooling is modeled using CLOUDY \citep{1998PASP..110..761F} for $T \gtrsim 10^{4}$ K. Below $10^{4}$ K we adopt the cooling rates from \cite{1995ApJ...440..634R}. We allow the gas to cool radiatively to 100 K, but adiabatic cooling can lower the temperature below this threshold. The UV background is derived from \cite{1996ApJ...461...20H}.

We have also modified \textsc{ramses} to include a simple molecular cooling model that is important for low-temperature cooling in the pristine gas \citep{2006MNRAS.366..247J, 2008arXiv0809.2786P, 2013ApJ...763...52H}. Our analytic model is based on \cite{1996ApJ...461..265M} and provides a radiative cooling rate, $\Lambda_r/n_{H2}$, per H$_{2}$ molecule across the range of densities encountered in the simulation. The details are found in \cite{2017ApJ...834...23S}. 

Star particles (SPs) are spawned in regions of gas according to a Schmidt law (Schmidt 1959) with 
\begin{equation}\label{eqn:sf}
\frac{d\rho_{\star}}{dt} =  \epsilon_{\star} \frac{\rho}{t_{\rm ff}} \theta(\rho- \rho_{\rm th}),
\end{equation}
where the Heaviside step function, $\theta(\rho- \rho_{\rm th})$, allows for star formation only when the density exceeds a threshold value $\rho_{\rm th}.$  We have set $\rho_{\rm th}$ to be the maximum of $1.0 \, m_p \, {\rm cm}^{-3}$ and 200 times the mean density in the simulation. These criteria ensure that SPs are only formed in virialized halos and not in high-density regions of the cosmological flow \citep{2006A&A...445....1R,2008A&A...477...79D}. We set the star forming efficiency to $\epsilon_{\star} = 0.01$, a value that results in reasonable agreement with the observed cosmic star formation rate \citep{2016PASA...33...37F, 2014ARA&A..52..415M}. The gas freefall time is $t_{\rm ff} = \sqrt{3 \pi /(32 G \rho)}$. 

Each SP models a \cite{1955ApJ...121..161S} (for polluted stars with $Z > Z_{\rm crit}$) and a log-normal (for Pop III stars) IMF. Our SP mass resolution is dictated by the star-forming density threshold and our resolution resulting in $m_{\star} = \rho_{\rm th} \Delta x^{3} = 6.6 \times 10^{3}\, M_{\odot}$. The final mass of each SP is drawn from a Poisson process such that it is a multiple of $m_{\star}$.

A fraction of the each SP's mass is returned to the gas in the form of supernovae (SNe). This occurs after the 10 Myr lifetimes for the most massive stars in the IMF \citep{2008ApJ...689..358R}. The impact of these SNe is parameterized by the fraction of the SP mass they eject, $\eta_{\rm SN}$, and the kinetic energy per unit mass of this ejecta, $E_{\rm SN}$. We take $\eta_{\rm SN}=0.10$ and $E_{\rm SN}=10^{51}$ ergs/10 $M_{\odot}$ for all stars formed throughout the simulation. The fraction of new metals in SN ejecta is 0.15 even though metal yields and energy from Pop III stars are likely to have been higher \citep{2003ApJ...589...35S,2005ApJ...624L...1S}. We may explore different yields and the subsequent effect on stellar enrichment in future work.  

We do not model radiative transfer or radiation pressure. While radiation pressure from massive young stars can disrupt star formation \citep{2012ApJ...745...50W, 2004ApJ...610...14W} it can also trigger it in dense clumps of gas \citep{2012A&A...546A..33T, 2010A&A...523A...6D}. While we have not modeled its effects for this work, it will be important to characterize the effects of radiative feedback in future work.

\subsection{The Pristine Fraction and the Corrected Metallicity}\label{sec:PF}
In order to more accurately model the fraction of Pop III stars created throughout cosmic time, we track two new metallicity-related quantities.  The \textit{pristine gas mass fraction}, $P$, models the mass fraction of gas with $Z < Z_{\rm crit}$ in each simulation cell. The evolution of this scalar tracks the time history of metal mixing within the cell such that when $P=0$ the entire cell has been polluted above $Z_{\rm crit}$. The scalar $P_{\star}$ records, for all time, the value of $P$ in star particles at the time they are spawned and indicates the mass fraction of the SP with $Z_{\star} < Z_{\rm crit}$. 

A simple equation can be used to describe the evolution of the pristine gas fraction in simulation cells:
\begin{align}\label{eq:selfConv}
\frac{d P}{d t} = - \frac{n}{\tau_{\rm con}}  P(1-P ^{1/n}). 
\end{align}
This equation traces the evolution of $P$ as a function of $n$ and a timescale $\tau_{\rm con}$, which, in turn, are functions of the turbulent Mach number, $M$, and the average metallicity of the cell relative to the critical metallicity, $\overline Z /Z_{\rm crit}$ \citep{2010ApJ...721.1765P, 2012JFM...700..459P, 2013ApJ...775..111P, 2017ApJ...834...23S}. Modeling the decay of the pristine gas fraction allows us to track the formation of Pop III stars as a mass fraction of all stars created, even in cells with an average metallicity above critical. 

Each SP in the simulation is tagged with the average metallicity of the medium from which it was born, $\overline Z \rightarrow \overline Z_{\star}$. Furthermore, by knowing the average metallicity, $\overline Z$ (or $\overline Z_{\star}$ for SPs), and the pristine gas fraction, $P$ ($P_{\star}$), we can better model the metallicity of the polluted fraction of gas (or stars). More explicitly, since $\overline Z$ represents the average metallicity of a parcel of gas, and the polluted fraction, $f_{\rm pol} \equiv 1-P$, models the fraction of gas that is currently polluted with metals, we can use the value of $f_{\rm pol}$ to predict the enhanced, or corrected, metallicity,
\begin{equation}\label{eq:zcorr}
\begin{aligned}
Z = \frac{\overline Z} {f_{\rm pol}},
\end{aligned}
\end{equation}
of the polluted fraction of gas in each simulation cell. Similarly, $Z_{\star}$ captures the corrected metallicity of SPs. As expected, when $f_{\rm pol} = 1$ the corrected metallicity is the average metallicity.

The metallicity of the polluted fraction as described by Eqn. (\ref{eq:zcorr}) is only precise when all of the metals are contained in the polluted fraction. This is true only in regions where the pristine gas is first polluted by Pop III SNe. However, it is possible for some of the metals to be distributed in the pristine gas fraction defined as $0 \le Z < Z_{\rm crit}$. As discussed in \cite{2017ApJ...834...23S}, this results in a small uncertainty in the resulting corrected metallicity of our SPs that we will ignore in this work. However, we can easily bound the correction to metallicity. While equation (\ref{eq:zcorr}) captures the upper bound, the lower bound on the correction is 
\begin{equation}\label{eq:lowerlim}
\begin{aligned}
Z = \frac{\overline Z - Z_{\mathbb {P}} P }{f_{\rm pol}},
\end{aligned}
\end{equation}
where $Z_{\mathbb {P}} = Z_{\rm crit} = 10^{-5} Z_{\odot}$ is the upper limit on the metallicity of the pristine gas. If the pristine fraction has $Z_{\mathbb {P}}=0$, as it would when polluting the primordial gas, we recover equation (\ref{eq:zcorr}).  Even when considering this uncertainty, the corrected metallicity, $Z$, allows us to more accurately model the metallicity of our gas and SPs than would be possible using the average metallicity alone. 

Lastly, we note that we do not create polluted stars when $f_{\rm pol} < 10^{-5}$. In this case, we assume that all stars formed in the cell are Pop III since only a tiny fraction of the cell is polluted with metals. While this may seem arbitrary, it is used for convenience as such a small fraction of Pop II stars does not detectably contribute to the luminosity of our galaxies over the entire redshift range analyzed.

%

\subsection{Halo Finding}
We use the AdaptaHOP halo finder by \cite{2004MNRAS.352..376A} to find star-forming regions in the simulation volume at each redshift of interest. Only halos with at least 100 DM particles, corresponding to a DM halo mass of $1.4 \times 10^{7}\, M_{\odot}$, are considered by AdaptaHOP. Groups of 20 particles are used to compute the local density of a candidate halo and only objects with a density 80 times the average total matter density are stored.

\begin{figure*}[!ht]
\begin{center}
\begin{tabular}{cc}
\includegraphics[width=.95\columnwidth]{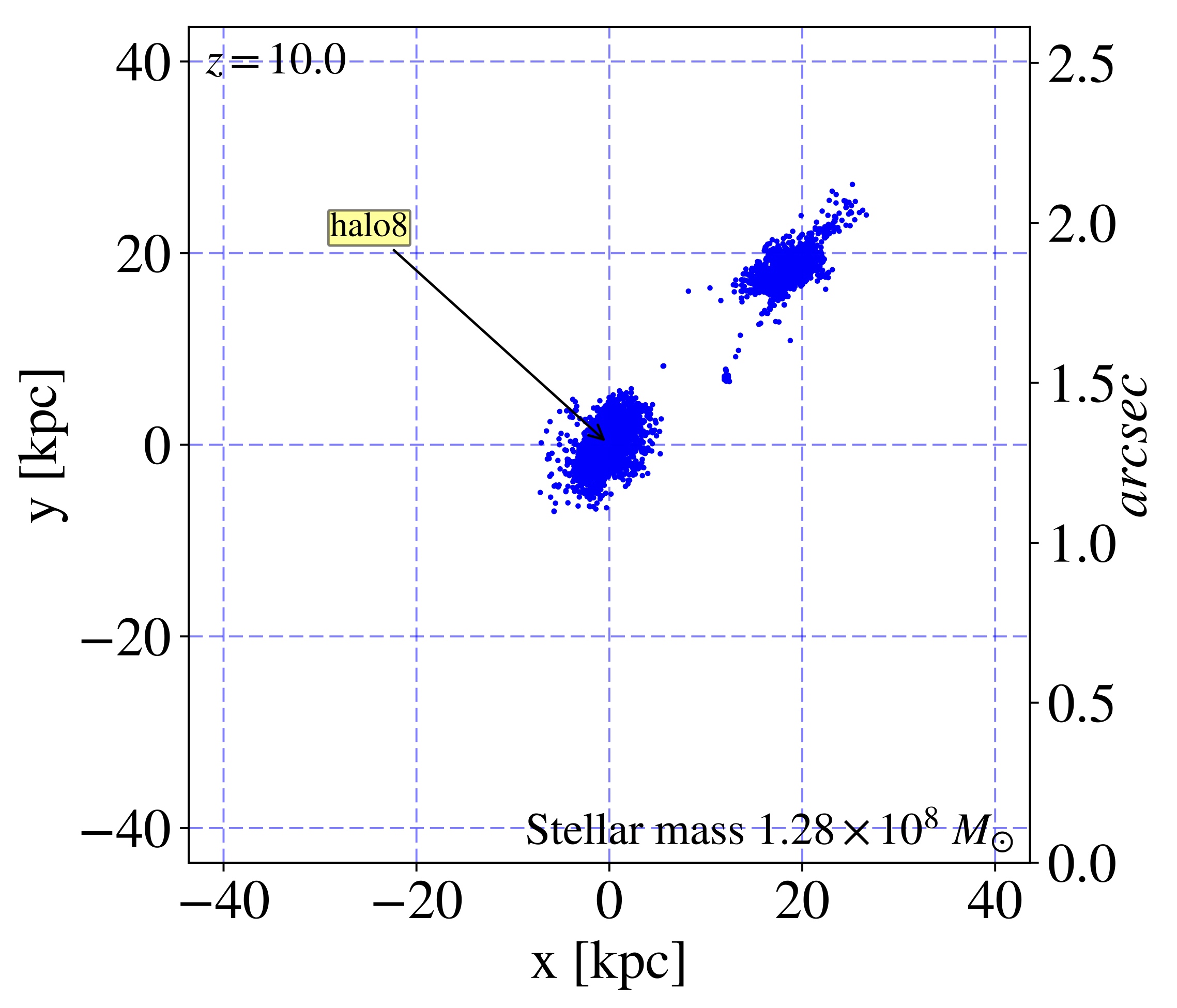} &
\includegraphics[width=.95\columnwidth]{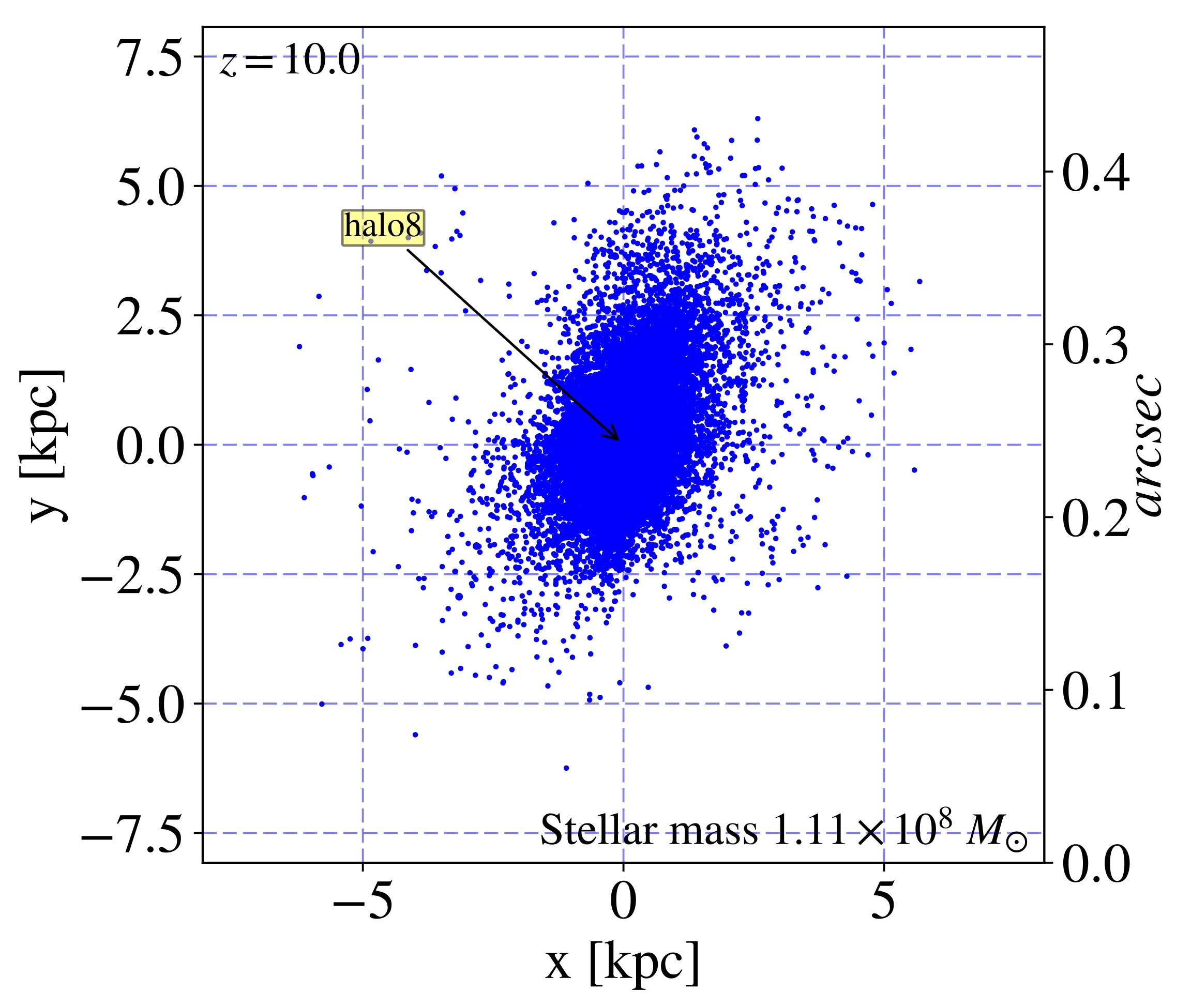} \\
\end{tabular}
\caption{Scatter plots of high-redshift galaxies in our simulation. Blue points are SP locations relative to the center of the located halo (coordinate 0,0). The image on the left depict the AdaptaHOP halo. As can be seen, more than one observationally identifiable galaxy is plotted. On the right, we have used our post-processing algorithm to correctly identify the larger of the two galaxies in the original field. The smaller galaxy was also identified but not depicted independently in this figure. The scale is comoving kpc and the total mass of the galaxy is identified in the lower-right of each plot. The scale on the right axes indicates the size of the field in arcseconds. Halo number 8 indicates that this halos was the 8th largest, by mass, at the redshifts indicated. }
\label{fig:galaxs}
\end{center}
\end{figure*}

Several of the more massive objects found by AdaptaHOP consist of more than one observationally distinguishable galaxy. Hence, we postprocessed the halos as follows. For each AdaptaHOP halo, we compute a mass, in stars, within a 3 kpc comoving sphere centered on the halo's coordinates. This typically corresponds to the core of the most massive galaxy in the field. Next, we iteratively compute the mass in larger concentric spheres about this core. At each step, we increase the radius by $10^{-1}$ arcsec converted to a proper distance (in kpc) at the galaxy's redshift. By using a redshift-dependent step size based on the observational reference frame, we can roughly determine the boundaries of our galaxies, assuming, as is possible with the HST, that objects on the order of 0.1 arcsec apart are distinguishable. We continue increasing the radius until the fractional change in enclosed mass is less than one part in $10^{4}$. Specifically, when $\nicefrac{\Delta M_{\rm enc,i}}{M_{\rm enc,i}} < 10^{-4}$, we consider the current radius to be the radius of a single galaxy. Figure~\ref{fig:galaxs} depict the galaxies associated with an unreprocessed AdaptaHOP halo (left) and the resolved galaxies (right) that result from using this procedure. The approach ensures we do not overrepresent bright objects by considering multiple galaxies as one when computing their luminosities.

To ensure that we capture the faint end of the LF, ignoring simulation resolution effects for now, we also locate and analyze the `missing' galaxies in our simulation, i.e., those that may have been missed by AdaptaHOP as configured. To accomplish this, we collect the locations of all SP at each redshift that are not within the previously computed radii of AdaptaHOP galaxies. This results in a set of temporarily orphaned SPs. Next, we select an SP from this orphan list and locate all SPs within a 2 kpc comoving radius. If there are none, we assume the star is a galactic outlier, ignore it for the current iteration, and select another SP. Given a collection of SPs within 2 kpc, we compute the center of mass of this set and use this new location with our expanding sphere method to find the extent of the galaxy. If the resulting object has $M_{\rm G} > 10{^4}\;M_{\odot}$, its center of mass location and radius are added to the list of galaxies and stored; otherwise, it is ignored. In either case, all of the object's SPs are then removed from the orphan list, and the procedure is repeated until all SPs have been processed. 
\subsection{Galaxy Spectral Models}
The rest-frame UV and filter fluxes of our simulated galaxies are functions of the ages, metallicities, and masses of their constituent SPs. We calculate our SP luminosities using a set of simple stellar population (SSP) SED models spanning the particles' ages and metallicity range. Our SEDs are based on \textit{STARBURST 99} \citep{2014ApJS..212...14L}, henceforth \textit{SB99} along with \cite{2010A&A...523A..64R} and \cite{2003A&A...397..527S}, henceforth \textit{R10}. For the fraction of all SPs with $Z_{\star} \geq Z_{\rm crit}$ our SEDs model a \cite{1955ApJ...121..161S} IMF normalized to $1\, M_{\odot}$. Since we have a precise age for each star particle, our SEDs model instantaneous bursts across the age range of SPs in the simulation. Pop III SP fractions with $Z_{\star} < Z_{\rm crit}$ are modeled using a log-normal IMF, again normalized to $1\, M_{\odot}$ and are based on the \textit{R10} SEDs for a zero-metallicity population. The log-normal IMF is centered on a characteristic mass of $60M_{\odot}$ with $\sigma=1.0$ and a mass range $1\, M_{\odot} \le M \le 500\, M_{\odot}$. Conceptually, Pop III stars include the mass of SPs with corrected metallicities $0 < Z_{\star} < Z_{\rm crit}$ as well as the fractional mass of pristine stars, $P_{\star}\times M_{\star}$, with $Z=0$, that represent the mass fraction of Pop III stars born in cells with incomplete mixing.  Since $P_{\star}$ captures the fraction of stellar mass with $Z_{\star} < Z_{\rm crit}$ the total mass of Pop III stars in each of our simulated galaxies is
\begin{equation}\label{eq:pop3mass}
\begin{aligned}
M_{\star, \rm III} = \sum_{n=1}^{N}{P_{\star, n} \; M_{\star, n}} ,
\end{aligned}
\end{equation}
where $N$ is the total number of SPs in a galaxy and $M_{\star, n}$ is the mass of each SP. 

Our \textit{SB99} SEDs were generated over an age range of 10 kyr to 0.78 Gyr, the age of the universe at $z=7$, in linearly spaced steps of 0.5 Myr. Each SED covers the wavelength range $91 - 1.6\times10^{6} \AA$. We generated SEDs for metallicities of 0.02, 0.2, 0.4, and 1.0 $Z_{\odot}$, for each age, using the \textit{SB99}-implemented Padova \citep{2000A&AS..141..371G} models that include stellar and nebular emission through the onset of the thermal pulse assymtotic giant branch phase of stellar evolution. We supplemented the \textit{SB99} model with a set of \textit{R10} models for stars with $Z=5\times10^{-4}$ and $5\times10^{-6}\, Z_{\odot}$. This allows us to interpolate over the range $Z_{\rm crit} \le Z_{\star} \le Z_{\odot}$. The Pop III SEDs, by \textit{R10}, are based on $Z=0$ and cover the age range 10 kyr to 1 Gyr in steps of 1 Myr. Again, the spectrum of all stars with $Z_{\star} < Z_{\rm crit}$ is modeled using this SED.




In order to compute the observational flux, we redshift each of our SEDs over the range $z$=7-16 applying Lyman forest and continuum absorption as described in~\cite{1995ApJ...441...18M}. 
This process, along with a spectral conversion from wavelength to frequency, transforms the rest-frame \textit{SB99} and \textit{R10} SEDs (erg/s/$\AA/M_{\odot}$) into observational fluxes (erg/s/Hz/cm$^2/M_{\odot}$) across the range (in redshift, age, and metallicity) of our SP. Equation (\ref{eq:LtoF}) describes this conversion from rest-frame luminosity to observational flux for objects at cosmological distances,
\begin{equation}\label{eq:LtoF}
\begin{aligned}
f(\nu,z) =  \frac{L_\nu(\nu_e)}{4 \pi D_L^2} (1+z) \, \mathcal{M}(\nu_{o}, z),
\end{aligned}
\end{equation}
where the $\nu_{o}$ and $\nu_{e}$ are in Hz and refer to the observed and emitted reference frames, respectively; $D_L$ is the luminosity distance; and $\mathcal{M}(\nu_{o}, z)$ is the~\cite{1995ApJ...441...18M} Lyman absorption function. We also generate the flux at a distance of 10 pc to facilitate the generation of absolute magnitudes. This is done by setting $z=0$, $D_L = 10$ pc and $\mathcal{M}(\nu_{o}, z)=1.0$ in equation (\ref{eq:LtoF}).

We then convolve these bolometric fluxes with the set of JWST and HST filters listed in Table~\ref{tab:filts}. We also compute the rest-frame UV flux at 1500$\AA$. The observational fluxes are computed as follows:
\begin{equation}\label{eq:filterFlux}
\begin{aligned}
\mathcal{F}(R,z)=\frac{\int_{-\infty}^\infty{f(\nu,z) R(\nu)\, \frac{d\nu}{\nu} }}{\int_{-\infty}^\infty{R(\nu)\, \frac{d\nu}{\nu} }},
\end{aligned}
\end{equation}
where $f(\nu,z)$ is the flux at redshift z, $R(\nu)$ is the filter response function, and $\mathcal{F}(R,z)$ is the resulting bandpass flux. For the rest-frame UV flux, the filter response function is the simply the Dirac delta function shifted to the observational UV wavelength, $\nu_{\rm UV} =\nicefrac{c}{(1+z) 1500\AA}$, resulting in $R(\nu)= \delta(\nu - \nu_{\rm UV})$ which simplifies equation (\ref{eq:filterFlux}) to  $\mathcal{F}(R,z)=f(\nu_{\rm UV},z)$. The result is a set of filter-flux tables that span the range of redshifts, ages, and metallicities for a normalized star of $1 M_{\odot}$ representing the Salpeter IMF, for $Z_{\star} \ge Z_{\rm crit}$, and the log-normal IMF for $Z_{\star} < Z_{\rm crit}$. This set of filter-flux tables for each redshift can be interpolated (in two dimensions) over the range of SP ages and metallicities found in the simulation. 


\begin{deluxetable}{c|l}
\tabletypesize{\footnotesize}
\tablecolumns{2} 
\tablecaption{\label{tab:filts}Filters Modeled in This Study} 
\tablehead{\colhead{System} & \colhead{Filter Names} } 
\startdata
JWST NIRCam & F150W,  F200W, F277W, F356W, F444W \\ \hline 
HST WFC3 & F125W, F160W  \\ \hline  
Rest-frame UV & 1500 $\AA$ \\
\enddata 
\end{deluxetable} 

\subsection{Simulated Observations}

We interpolate the filter and rest-frame UV fluxes linearly in log-space as a function of both SP metallicity and age in order to compute the bandpass and rest-frame UV flux of our galaxies at each redshift. 
The resulting fluxes are then scaled by the mass of each SP, accounting for $P_{\star}$, and summed to compute the total flux (in each filter) for the galaxy. We then transform the filter fluxes into AB magnitudes.

\section{Results}
\begin{figure}[h]
\begin{center}
\includegraphics[width=1\columnwidth]{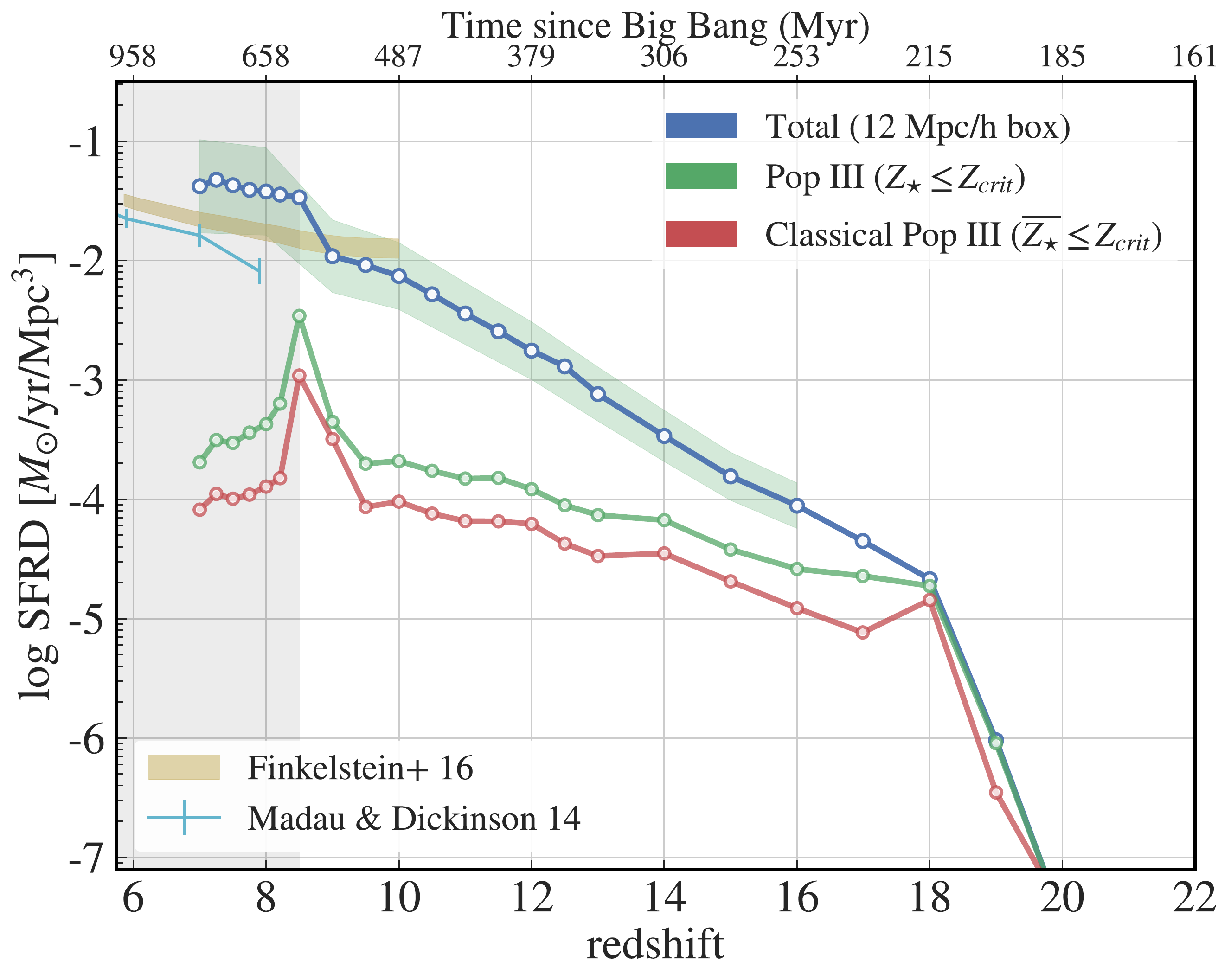}
\caption{Our SFRD along with observations by \cite{2014ARA&A..52..415M} and a LF-based SFRD by \cite{2016PASA...33...37F}. The 1$\sigma$ uncertainty on the overall SFRD is also depicted  (light-blue) out to $z=16$. While our SFRD is above observations at $z\le 8.5$, it agrees with the LF-based SFRD that incorporates galaxies down to $M_{\rm UV} = -13$ mag when considering both Poisson and sample variance, 1$\sigma$. The grey-shaded area indicates redshifts post-reionization. }
\label{fig:sfrd}
\end{center}
\end{figure}

In this section, we present the characteristics of our simulated galaxies. We focus on $7 \le z \le 15$. Figure \ref{fig:sfrd} depicts the star formation rate density (SFRD) for our simulation, along with an observationally derived SFRD from \cite{2014ARA&A..52..415M}. While our SFRD is higher than observations at $7 \le z \le 8$ it agrees with the LF-based SFRD described by \cite{2016PASA...33...37F} when considering sample variance \citep[][see Section $\ref{ee}$ for a discussion on error estimation]{2008ApJ...676..767T}. The LF-based SFRD is based on an integration of the reference luminosity function in that work to $M_{\rm UV}$ = -13 mag. Since the observationally based SFRD is likely undersampled at $z>7$ \citep{2015ApJ...808..104O}, the LF-based SFRD is likely a more appropriate estimate of star formation at high redshift. 

The figure also depicts the Pop III SFRD as well as (what we call) the ``classical'' Pop III SRFD that does not include the effects of modeling the evolution of $P$. We see that modeling the pristine fraction increases the SFRD for Pop III stars by an average factor of 2.5 for $z \le 16$. As we discuss below, a relatively small increase in the fraction of young Pop III stars can have a significant impact on the luminosity of early galaxies. 

There is a rapid increase in the star formation rate immediately before reionization ($z_{reion}\approx 8.5$) that correlates with an even greater increase in the Pop III star formation rate. This is caused by a significant number of new, small halos crossing the density threshold for star formation.  At $z=9$ the number of star forming halos is 2112. By $z=8$ that number rises to 6807, more than a factor of 3 increase in $\approx$96 Myr. While the overall star formation rate rises by a factor of approximately 3.5 from $z=9$ to $z=8$, the Pop III rate increases by a factor of 4.4 over this same interval. Additionally, the fraction of Pop III stellar mass in our simulation box at $z=8.5$ increases to 7\% from 4\% at $z=9$. 

\subsection{The Galaxy Mass-metallicity Relation}
\begin{figure*}[t]
\begin{center}
\includegraphics[width=\textwidth]{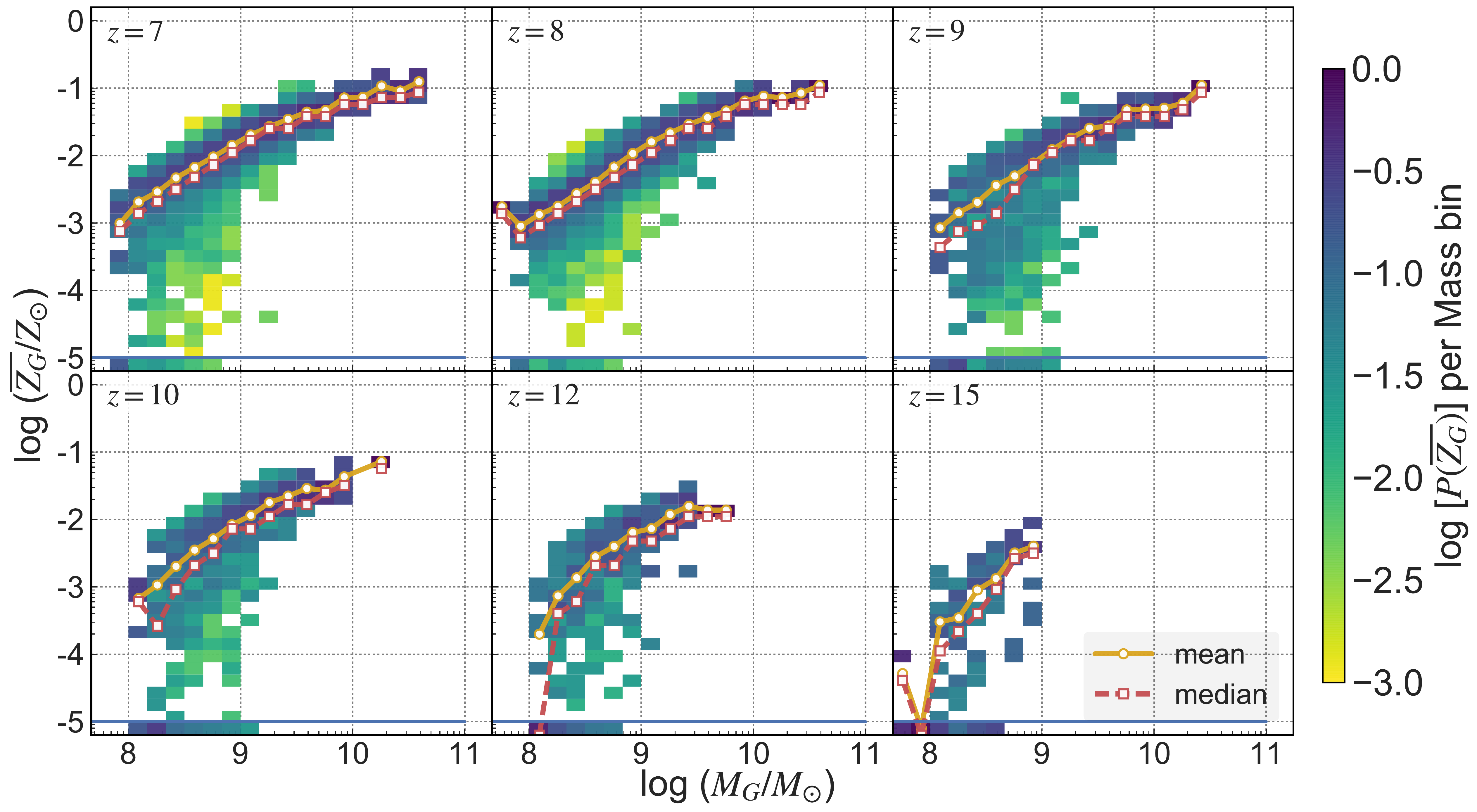}
\caption{The normalized probability, in each mass bin, of finding a galaxy with a metallicity in the range $1 > Z_{\rm G} \ge 9\times 10^{-6} Z_{\odot}$ where we have binned all Pop III galaxies immediately below $Z_{\rm crit} = 10^{-5} Z_{\odot}$ (blue line). The dark yellow line identifies the mean metallicity and the red dashed-line the median. Pop III galaxies with $M_{\rm G} > 10^{9} M_{\odot}$ are exceedingly rare within our simulation volume ($\approx$ 4800 Mpc$^3$, comoving) occurring only in the range $9 \le z \le 10$. At $z=8$ we see a large number of small Pop III galaxies that formed immediately before reionization with masses less than $10^{8} M_{\odot}$.  Note that $M_{\rm G}$ is the  halo DM mass. The typical galaxy stellar masses corresponding to the  halo DM masses are between 2.6 dex, at $z=7$ to 3.2 dex, at $z=15$ lower. }
\label{fig:gMZ}
\end{center}
\end{figure*}

While the galaxy mass-metallicity relation at $z \ge 7$ is beyond current observational limits \citep{2013ApJ...771L..19Z,2013ApJ...776L..27H,2008A&A...488..463M}, Figure \ref{fig:gMZ} depicts this relationship for a sample of redshifts in the range $7 \le z \le 15$ for our simulated galaxies. The plots display the normalized probability per mass-bin, $\sum_{\rm bin}{\nicefrac{P(\overline{Z_{\rm G}} / Z_{\odot})}{d(M_{DM}/M_{\odot})}} = 1.0$, of finding a galaxy with a metallicity in the range depicted on the vertical axis. Here, we use the halo DM mass for the galaxies. The figure clearly depicts the expected mass-metallicity trend but, more importantly for this work, the mass range of Pop III galaxies in the bottom row of bins in each plot, at each redshift. Each galaxy's average metallicity, $\overline{Z_{\rm G}}$, is computed using the corrected SP metallicities described by Equation (\ref{eq:zcorr}). Note that $\overline{Z_{\rm G}} $ is computed directly from the mass-weighted average metallicity of the SPs that populate each galaxy and not from synthetic observations of galaxy spectra. Pop III galaxies, composed  of SPs such that the average metallicity of the galaxy is subcritical, have been grouped at $\overline{Z_{\rm G}} < 10^{-5} Z_{\odot}$. We analyze halos with masses down to $M_{\rm G} = 4.62 \times 10^{7}M_{\odot}$ that consist of approximately 330 DM particles. 

Taken as whole, we see that Pop III galaxies are not very massive and are comparable to the theoretical limits, $ 1.5\times10^{8}M_{\odot}$ to $1.1\times10^{9}M_{\odot}$, discussed in \cite{2017MNRAS.467L..51Y} for $z=7$. The most massive Pop III-dominated galaxies in our simulation occur at $z=9$ and 10, before reionization. They have an average DM mass of $\overline{M_{\rm G}} = 1.2 \times 10^{9}M_{\odot}$ and make up less than 3\% of all galaxies with masses $M_{\rm G} > 10^{9}M_{\odot}$. 

At lower redshift, $z=7$ and 8, Pop III galaxies span a smaller mass-range where the most massive, less than 1\% of all Pop III galaxies, have $M_{\rm G} \gtrsim 4.6\times 10^{8}M_{\odot}$. At the other end of the mass range, we see the recently formed, purely Pop III galaxies with $M_{\rm G} < 10^{8}M_{\odot}$. At $z=7$ and 8 fully 69\% and 54\% of Pop III galaxies, respectively, are associated with these mini-halos. This is likely because the rate and location of Pop III star formation has changed between $z=9$ and $z=8$. The Pop III SFRD turns over at $z_{\rm reion}=8.5$ and the Pop III fraction is no longer keeping pace with overall star formation. 

While the majority of new star formation is taking place within larger, shielded galaxies -- and within gas that has been polluted to levels above $Z_{\rm crit}$ -- we also see the results of the Pop III starbursts in new, mini-halos immediately before $z=8.5$. 

The low masses of purely Pop III protogalaxies in the range $8\le z \le 11$, today's high-redshift frontier, partially explain the difficulty in finding Pop III galaxies. However, as we shall discuss, a small percentage of young Pop III stars can contribute a significant fraction of a galaxy's flux. 

\subsection{Error Estimation}\label{ee}
We briefly describe the error estimation for both the luminosity functions and for the overall SFRD. Error estimates include both Poisson errors (shot noise) and the 1$\sigma$ uncertainty in galaxy counts due to sample variance \citep{2008ApJ...676..767T} and are computed per luminosity bin. For the SFRD the process is the same except all galaxies are essentially in one bin per redshift.

The total error in each bin is
\begin{equation}\label{eq:toterr}
\begin{aligned}
v_r = \sqrt{\sigma_{v}^2 + 1/N},
\end{aligned}
\end{equation}
where the sample variance
\begin{equation}\label{eq:samperr}
\begin{aligned}
\sigma_{v} ^2= (\overline{b})^2\,\sigma_{\rm box}(z)^2
\end{aligned}
\end{equation}
is the product of the average galaxy bias, $\overline{b}$, based on \cite{1974ApJ...187..425P}, and the fluctuation amplitude, $\sigma_{\rm box}(z)$, for the simulation volume at redshift $z$. The shot noise is $1/N$. 

In turn, the average bias is derived from the mass of each galaxy in the bin,
\begin{equation}\label{eq:bias}
\begin{aligned}
b = 1 + \frac{(\nu^2-1)}{1.69},
\end{aligned}
\end{equation}
where
\begin{equation}\label{eq:nu}
\begin{aligned}
\nu = \frac{1.69}{ \sigma(M,z)}
\end{aligned}
\end{equation}
and $\sigma(M,z)$ is the fluctuation amplitude of a galaxy of mass M at redshift $z$. 

Lastly, the DM mass in collapsed objects, at each redshift,  matches the  prediction in \cite{2001PhR...349..125B} to within -3\% to +6\% at $z \le 10$. The greatest difference is at $z=12$ to 15 where the simulation has 12 to 14\% more mass in halos than predicted  by theory resulting in a slight overestimate of the sample variance at $z\ge 12$.

\subsection{Luminosity Functions}
Galaxy observations are characterized by their flux -- which, in turn, is determined by the galaxy's stellar populations. A small fraction of hot, young Pop III stars can contribute a large fraction of the galaxy's luminosity. However, only the Pop III stars with ages $< 3.5$ Myr contribute more flux than their polluted cousins, so detecting a galaxy dominated by Pop III flux means looking for a recent starburst such that a significant fraction of the flux from the entire galaxy is coming from these types of stars. We next look at the LFs and Pop III flux fractions derived from our simulation data. 

\begin{figure*}[!ht]
\begin{center}
\includegraphics[width=2\columnwidth]{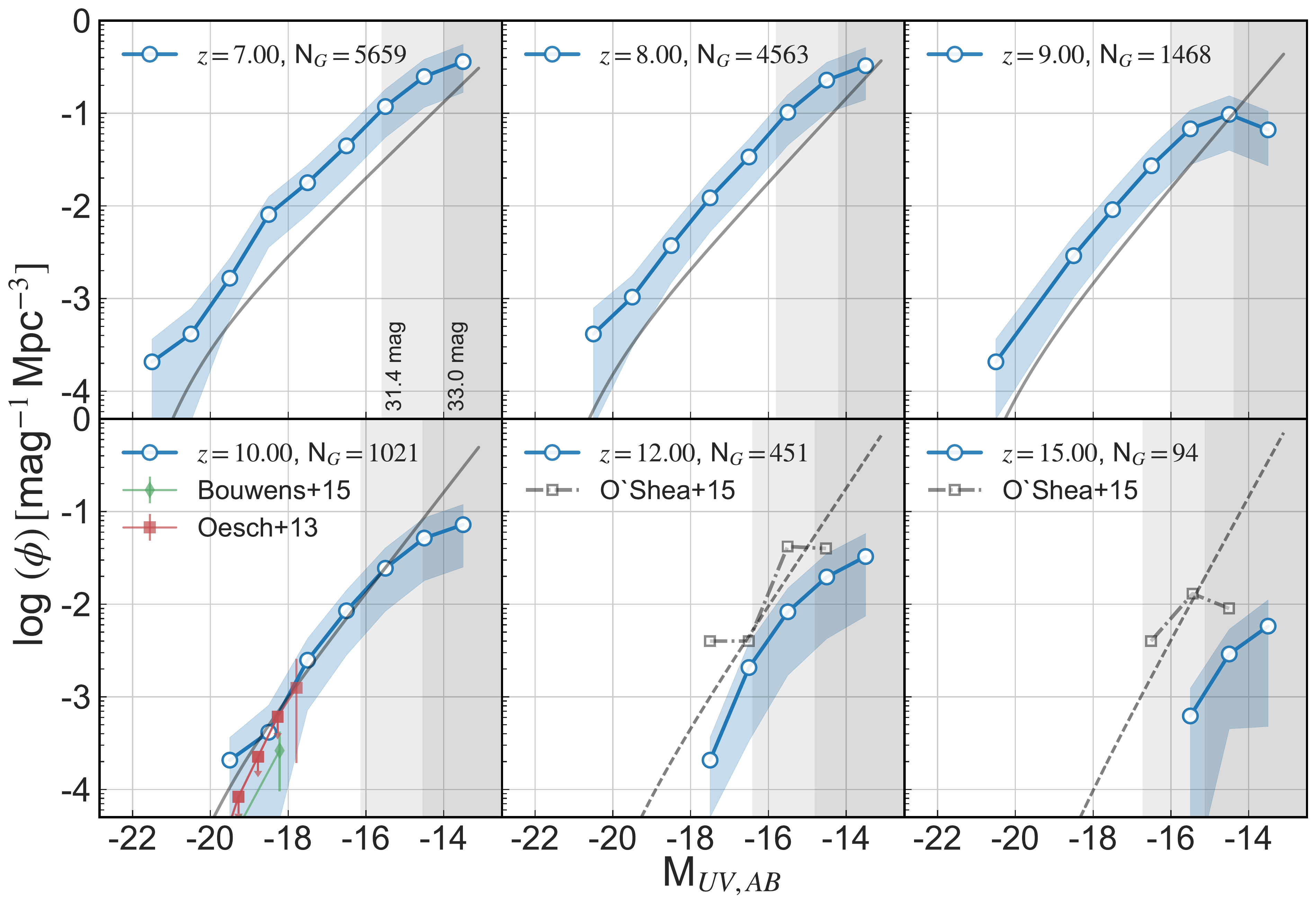}
\caption{UV luminosity functions derived from our simulation with 1$\sigma$ error bounds including both Poisson noise and sample variance. Dark grey lines are \cite{2016PASA...33...37F} Schechter fits. Dashed-grey lines are Schechter functions based on an extrapolation of the Schechter parameters also found in that work. For $z=10$, we have included \cite{2015ApJ...803...34B} and \cite{2013ApJ...773...75O} points based on observations, with error bars. For redshifts 12 and 15 we have included luminosity functions derived from the Renaissance Simulations by \cite{2015ApJ...807L..12O}. The shaded areas indicate the regions where $m_{\rm UV} > 31.4$ mag, a likely limiting magnitude for a JWST ultra-deep campaign and $m_{\rm UV} > 33$ mag, a likely lensing limit.}
\label{fig:fit}
\end{center}
\end{figure*}

Given our total simulation volume of 4828 Mpc$^{3}$ we have data down to $\phi \approx 2\times10^{-4}\, \,{\rm mag}^{-1}$ Mpc$^{-3}$.  Further, since star formation in our simulation is resolution-dependent we cannot track galaxy formation at scales below $\approx 260$ pc, physical. While such a small protogalaxy is likely not detectable, even by the JWST, it does prevent us from characterizing the turnover at the faint end of the LF. Additionally, several such mini halos may merge producing larger numbers of fainter galaxies than reported here. Within this context, Figure \ref{fig:fit} depicts the UV luminosity functions for all of our galaxies down to $M_{\rm UV} = -13$ mag, where the galaxy counts per magnitude bin begin to decrease due to the simulation's limited resolution. 

We have included both observationally-derived and extrapolated \cite{1976ApJ...203..297S} functions by \cite{2016PASA...33...37F} for reference: Solid grey lines indicate Schechter functions derived from observations, while grey-dashed lines are an extrapolation of the Schechter parameters -- also from \cite{2016PASA...33...37F}.  The Schechter parameters for the observational data and extrapolations are listed in Table \ref{tab:schecParams}. We have also included observational data from \cite{2015ApJ...803...34B} and \cite{2013ApJ...773...75O} along with data from an analysis of galaxies in the Renaissance Simulations by \cite{2015ApJ...807L..12O}. 

\begin{deluxetable}{r|p{0.2\columnwidth}p{0.2\columnwidth}p{0.2\columnwidth}}[h!]
\tablewidth{\columnwidth}
\tablecolumns{4}
\tablewidth{100pt} 
\tablecaption{\label{tab:schecParams}Schechter Function Parameters} 
\tablehead{ z & log($\phi^{*}$) &  $\alpha$ &  $M^{*}_{\rm UV}$ }
\startdata
    8 & -3.75 & -2.13 & -20.52 \\
    9 & -3.94 & -2.24 & -20.39 \\
    10 & -4.13 & -2.35 & -20.25 \\ 
    11 & -4.29 & -2.47 & -20.11 \\ 
    12 & -4.49 & -2.58 & -19.98 \\
    13 & -4.69 & -2.69 & -19.84 \\
    14 & -4.89 & -2.81 & -19.71 \\
    15 & -5.08 & -2.92 & -19.57 \\
    16 & -5.28 & -2.03 & -19.44 \\
\enddata  
\tablecomments{Schechter function parameters for the reference lines in the luminosity function plots. Data is from \cite{2016PASA...33...37F}. Values at $z > 10$ have been extrapolated based on a linear fit to the parameters in that work.}
\end{deluxetable}

We note that all of our LFs at $z<10$ lie slightly above the faint end of observationally derived Schechter functions. However, it should be noted that data from our simulation includes many faint objects for which the observationally derived models suffer from the greatest uncertainty. Furthermore, although our simulation box represents an average density region of the universe, the variance in initial conditions could have resulted in more overdensities with a scale that is responsible for the increase in star formation at $z\approx9$. 

Additionally, we do not account for dust. Dust attenuation in high-redshift galaxies is uncertain at best \citep{2017MNRAS.470.3006C, 2017arXiv170202146C,2001PASP..113.1449C} and we have not included its effects in any of our plots. However, if we extrapolate work by \cite{2015A&A...574A..19S} at $z\approx 6.8-7.5$ to $z=8-10$, we would expect $A_{\rm UV}\approx 1.1\pm{0.2}$ of UV dust attenuation. Including this level of dust attenuation would reduce our absolute magnitudes by $\approx 1$ and bring our data more in-line with the faint end slope at these redshifts.  

Our LFs closely follow the predicted faint end slope, $\alpha$, at $z=10$ and are in reasonable agreement with both the extrapolated Schechter function and data from the Renaissance Simulations at $z=12$. Again, these Schechter curves (grey dashed lines) are based on a linear fit and extrapolation of the trends in $M^{*}$, $\alpha$, and log $\phi^{*}$ using observational data over the range $4\le z \le 8$.  Although we have no data at the bright end of the LF, due to our small volume, we feel that our LFs are reasonably representative of galaxy populations, in the range plotted, for an average-density region of the universe.

\subsection{Pop III Flux}
\begin{figure*}[h!tb]
\begin{center}
\includegraphics[width=2\columnwidth]{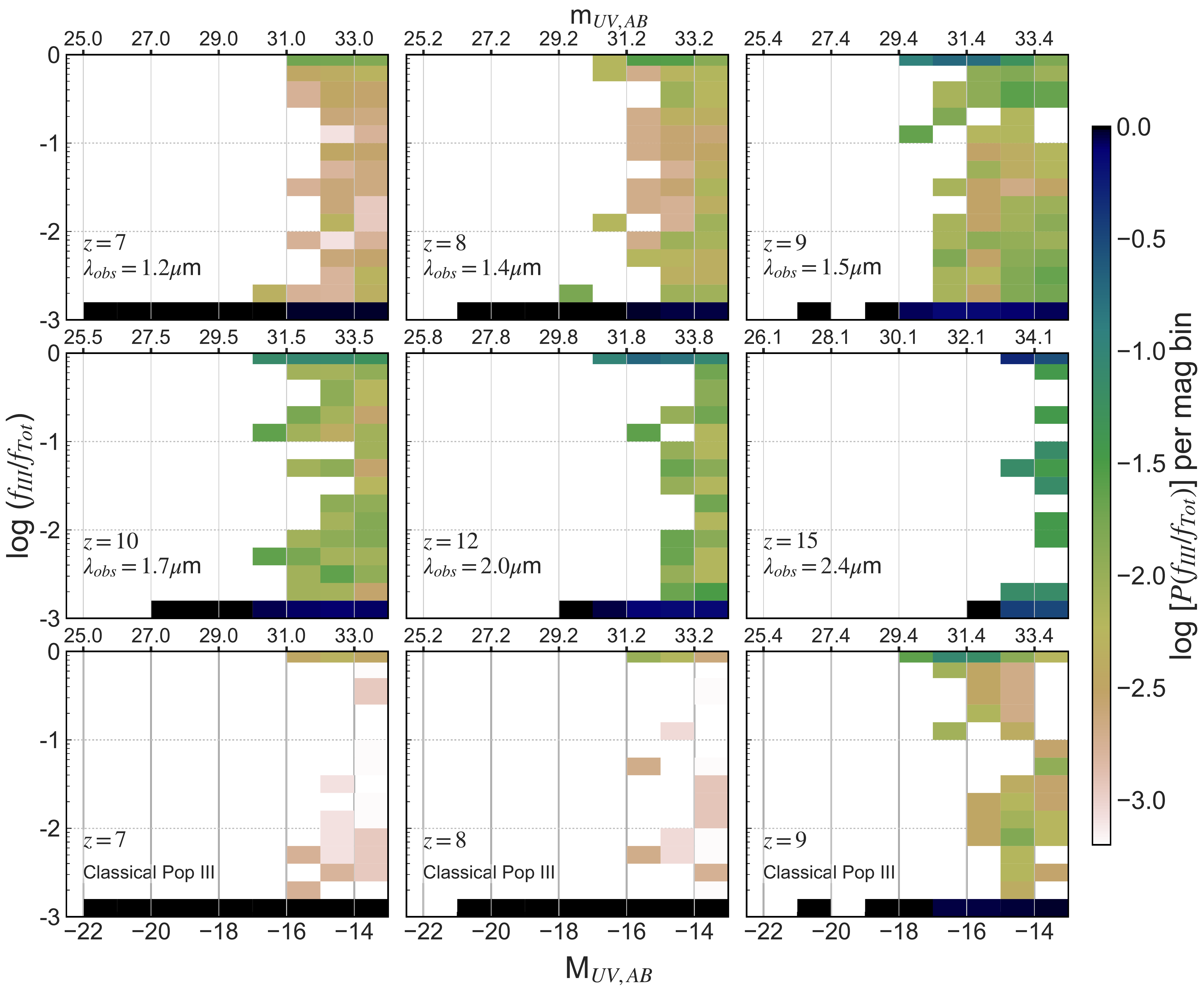}
\caption{The normalized probability of finding a UV Pop III flux fraction, $\nicefrac{f_{\rm III}}{f_{\rm Tot}}$, as a function of the redshift and magnitude of our galaxies.  When $\nicefrac{f_{\rm III}}{f_{\rm Tot}} < 10^{-3}$ we map the value to $10^{-3}$. Probabilities are computed independently for each magnitude bin. Bins are labeled at their right edge hence the far right bin is $M_{\rm UV} = -13$. The top-most row of bins in each plot represent a Pop III flux fraction of at least 75\%: $\nicefrac{f_{\rm III}}{f_{\rm Tot}} \ge 0.75$. The second row of bins represent $.75 > \nicefrac{f_{\rm III}}{f_{\rm Tot}} \ge 0.50$. At $z=9$, we find that 30\% of galaxies at $M_{\rm UV} \le -16$ mag have $\nicefrac{f_{\rm III}}{f_{\rm Tot}}> 75\%$.  The bottom row of plots depict the Pop III flux fraction from our galaxies when only considering stars created in cells with $\overline{Z} < Z_{\rm crit}$, the classical Pop III case.  Modeling the evolution of the pristine gas faction at subgrid scales results in a Pop III SRFD that is a factor of 2.5 increase over the classical rate and these luminous stars contribute a significant fraction of the flux of these young galaxies. Axis labels along the top axis are observed UV magnitude, $m_{\rm UV}$.  We identify $\lambda_{obs}$ at each redshift: the wavelength of the 1500$\AA$ reference in the observational frame. }
\label{fig:hist9}
\end{center}
\end{figure*}

Since we are mainly concerned with the search for Pop III stars, we focus our analysis on more detailed characteristics of our galaxies. Figure~\ref{fig:hist9} depicts the normalized probability of finding a Pop III flux fraction, as measured at $1500 \AA$ in the rest-frame, in the range $10^{-3} \le \nicefrac{f_{\rm III}}{f_{\rm Tot}} \le 1$ for our galaxies as a function of magnitude and redshift. When $\nicefrac{f_{\rm III}}{f_{\rm Tot}} < 10^{-3}$ we have mapped the value to $10^{-3}$. Note that probabilities are computed independently for each magnitude bin, as was done for the galaxy mass-metallicity relation. 

The topmost row of bins in each plot represent a Pop III flux fraction of at least 75\%: $P(\nicefrac{f_{\rm III}}{f_{\rm Tot}} \ge 0.75)$, while the next row down indicates a flux fraction $P(0.75 > \nicefrac{f_{\rm III}}{f_{\rm Tot}} \ge 0.50)$.  Note that combining the probabilities in the 50\% and 75\% bins does not change the probabilities significantly from considering the 75\% bins alone.  Hence we use $75\%$ as our definition of ``significant Pop III flux'' and a ``Pop III--bright galaxy''. Magnitude bins are labeled at their right edge and are 1 magnitude wide. Below we reference a magnitude bin by its right (dimmer) edge.

At redshift 7, only 2\% of galaxies with binned absolute magnitudes of -13, -14 and -15 are Pop III--bright. Similarly, at $z=8$ less than 1\% of galaxies are Pop III--bright and have $M_{\rm UV} = -16$ mag. However, as we move to the era before reionization, approximately 18\% of our galaxies at $z=9$ with $M_{\rm UV} = -15$ mag (corresponding to $m_{\rm UV} \approx 31.4$ mag at this redshift) are Pop III--bright and 11\% have $M_{\rm UV} = -17$ mag. This correlates with our observation of the increase in the SFRD at this epoch. At $z=10$, we find that the fraction of Pop III--bright galaxies drops to $\approx$ 8\% with $M_{\rm UV} = -15$ mag and 7\% with $M_{\rm UV} = -16$ mag. As we move to $z=12$, about 10\% of the faint objects ($M_{\rm UV} = -16$ mag) are dominated by Pop III flux. At $z=15$ the brightest Pop III--bright galaxies have $M_{\rm UV} = -14$ mag but represent 50\% of galaxies at that absolute magnitude. 

The results discussed so far include Pop III stars created in cells in which the subgrid turbulent mixing of metals was incomplete, resulting in the enhanced Pop III SFRD we see in Figure~\ref{fig:sfrd}. The bottom row of Figure \ref{fig:hist9} depicts the Pop III flux fraction for our galaxies when constraining Pop III star formation to cells with $\overline{Z} < Z_{\rm crit}$.  This is the no-mixing or classical Pop III case. When considering only the classical Pop III SPs we see that the enhancement of the Pop III SFRD due to our subgrid turbulent mixing model, an average of $\approx 2.5 \times$ the classical rate, is responsible for a significant amount of flux at several redshifts. 

For instance, considering all `classical Pop III galaxies' at $z=9$, only 7\% of galaxies with $M_{\rm UV} = -15$ mag are Pop III--bright, as compared to the 18\% we discuss above when we consider Pop III stars created in regions of incomplete mixing. The subgrid model results in $\approx$ 2.6 times more Pop III bright galaxies at this redshift and absolute magnitude.  This result points to the importance of accurately modeling Pop III star formation since small changes in their density can significantly effect the predicted fraction of Pop III flux.  

\begin{figure}[h!tb]
\begin{center}
\includegraphics[width=\columnwidth]{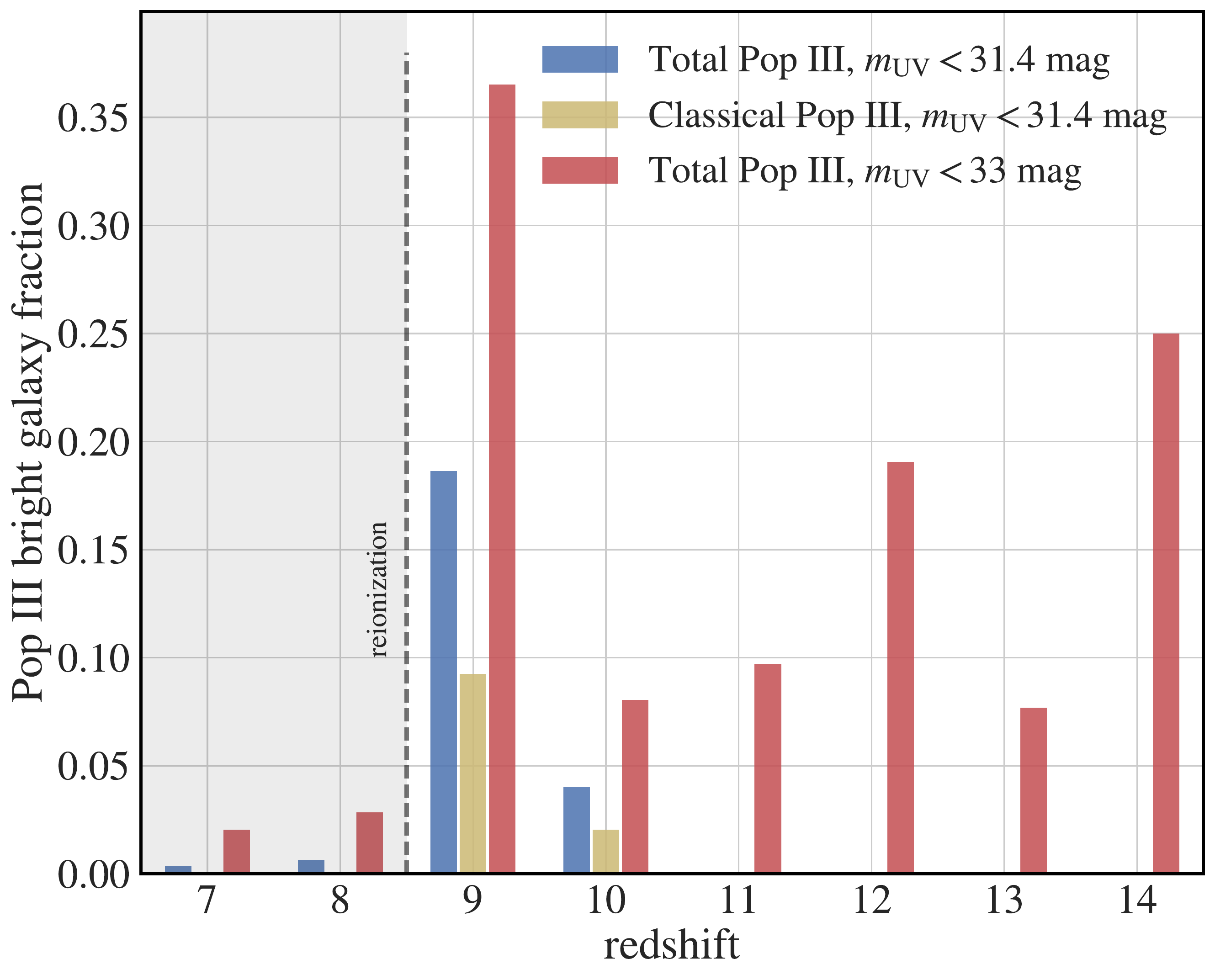}
\caption{The blue bars indicate the joint probability of finding an observable ($m_{\rm UV} \le 31.4$ mag) Pop III--bright galaxy as a fraction of all observable galaxies, as a function of redshift. The yellow bars ignore the effects of our subgrid model and its enhancement to the Pop III SFRD. The resulting reduction in Pop III flux is visible across  $7 \le z < 11$. The red bars also depict this joint probability but for a limiting intrinsic magnitude of $m_{\rm UV} = 33$ mag.  The burst of Pop III star formation immediately before reionization is apparent at $z=9$.}
\label{fig:Pop3Gal}
\end{center}
\end{figure}

Next, we consider the overall fraction of observable galaxies in the simulation that are Pop III--bright, at each redshift. Figure~\ref{fig:Pop3Gal} identifies the joint probability that a galaxy has at least a 75\% Pop III flux fraction and $m_{\rm UV} \le 31.4$ mag, which we take as the limiting magnitude for the un-lensed JWST ultra-deep campaign, as a fraction of all galaxies with $m_{\rm UV} \le 31.4$ mag. We refer to these galaxies as ``observable Pop III--bright galaxies''. As we would expect from current surveys, at relatively low redshift, $7\le z \le 8$, the fraction of Pop III--bright galaxies is less than 2\%. 

Going deeper, we again see the relatively large increase in the number of Pop III--bright galaxies at $z=9$, immediately after a burst of Pop III star formation, where 17\% of observable galaxies are Pop III--bright. This is the epoch immediately before reionization when smaller mini-halos begin to cross the star-forming mass-density threshold. It is during this epoch that we predict the largest fraction of detectable Pop III--bright galaxies. 

After reionization, the  star-forming threshold is raised quenching star formation in these mini-halos. This result points to the importance of determining the reionization redshift since most Pop III--bright galaxies are likely to be found just before it completes. 

At $z=10$ we note that only 5\% of our observable galaxies are Pop III--bright. At $z> 10$ there are no Pop III--bright galaxies with $m_{\rm UV} \le 31.4$ mag. To find Pop III--bright galaxies we have to go to $m_{\rm UV} = 33$ mag, an intrinsic magnitude that may be within reach of a lensed JWST field.

To once again illustrate the observational effects of our subgrid model, Figure~\ref{fig:Pop3Gal} also identifies the fraction of observable Pop III--bright galaxies when we only account for classical Pop III stars created in simulation cells with $\overline{Z} < Z_{\rm crit}$. As can be seen the subgrid models' resulting enhancement to galactic Pop III flux is evident over the redshift range $7 \le z < 11$. Comparing results in the redshift range $9 \le z \le 10$ we note that the fraction of observable Pop III--bright galaxies is, on average, 2x higher for our subgrid model than for the classical Pop III case. Again, this exemplifies the importance of modeling Pop III star formation accurately since it has a significant effect on the density of Pop III--bright galaxies we expect to detect at high redshift.  

Most of the Pop III--bright galaxies form at the border of polluted areas or in regions of pristine gas away from larger halos. While our sample volume is relatively small, this result points out that Pop III--bright galaxies can be found both in relative isolation and near other, often larger galaxies with $\overline Z_{\rm G} > Z_{\rm crit}$. Once again, modeling the mixing time required to pollute the gas above $Z_{\rm crit}$ is important here.

By examining fainter galaxies we can find a larger fractions of galaxies with significant Pop III flux at higher redshift. Figure \ref{fig:Pop3Gal} also depicts characteristics of galaxies that have at least 75\% of their flux coming from Pop III stars while requiring that $m_{\rm UV} \le 33$ mag, approximately the JWST 10x lensing limiting magnitude. With these criteria we note that at $z=11$ the fraction of Pop III--bright galaxies is only $\approx$ 10\%, the result of more galaxies dominated by Pop II flux meeting the criteria $m_{\rm UV} \le 33$ mag. However, at $z=12$ the fraction of observable Pop III--bright galaxies jumps to 19\% as a result of going to this intrinsic magnitude with lensing. At $z=14$ 25\% of galaxies are Pop III--bright. If there are enough lensing opportunities JWST should detect a reasonable (more than one in ten) fraction of Pop III--bright galaxies at $z=14$.  

\subsection{Observational Predictions}
In this section we discuss predictions for the space telescopes and filters described in Table \ref{tab:filts}. As with the rest-frame UV flux, we have not modeled dust for the results presented in this section. 

\begin{figure*}[h]
\begin{center}
\includegraphics[width=2\columnwidth]{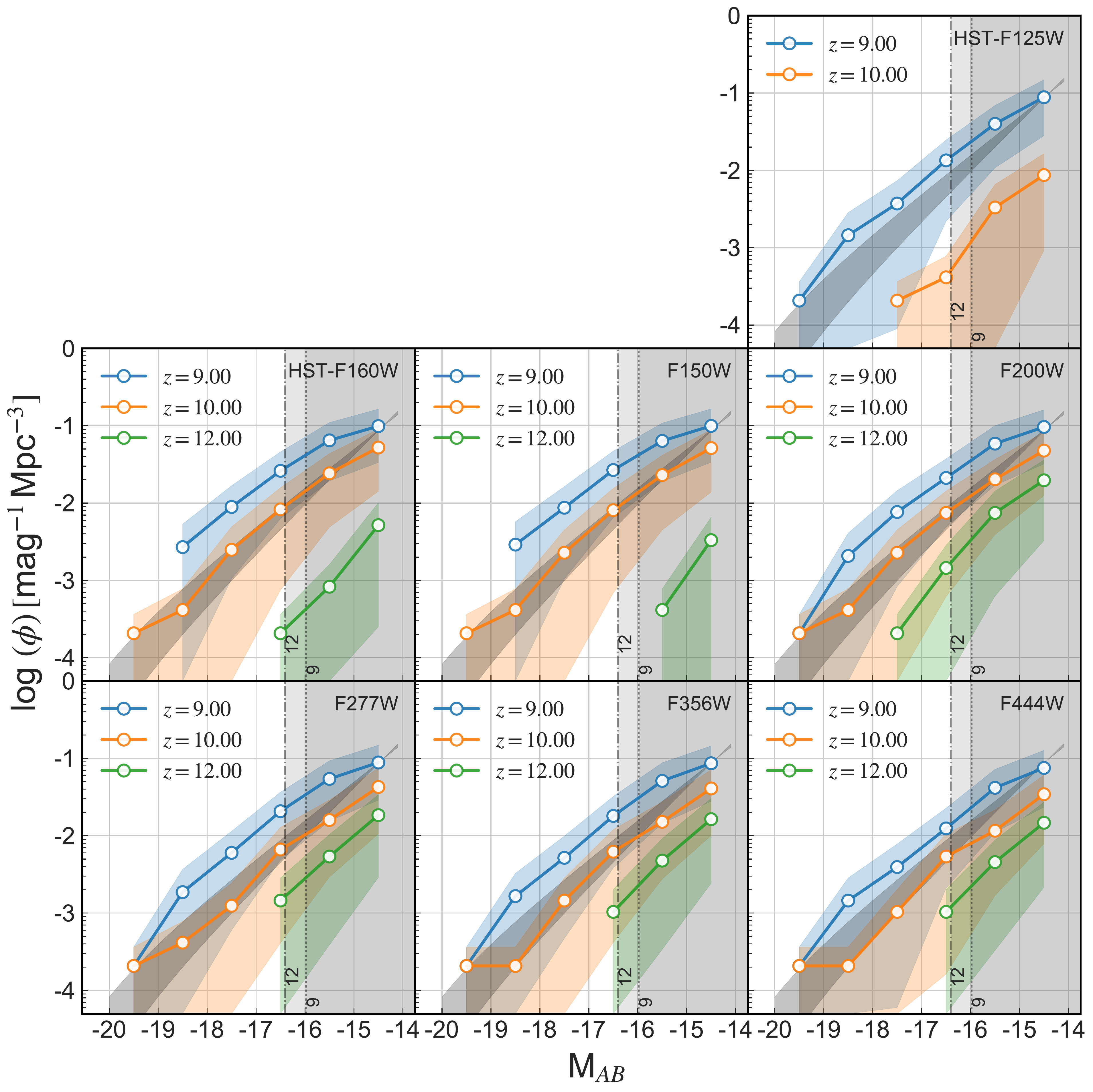}
\caption{Luminosity functions, with 1$\sigma$ error bounds, derived from our simulated galaxies convolved with our filter models across the redshift range $9 \le z \le 12$. The dark grey Schechter functions represent the bounding redshifts and are again from \cite{2016PASA...33...37F} (without errors). The dark, vertical shaded areas of each plot indicate the regions where $m_{\rm AB} > 31.4$ mag, the JWST limiting magnitude for the ultra-deep campaign, for $z=9$ and $z=12$. If a redshift does not appear in a plot, none of our galaxies were visible in that filter. Note that we have not included dust attenuation.}
\label{fig:filtLF1}
\end{center}
\end{figure*}

The LFs derived from our simulated bandpasses are depicted in Figure \ref{fig:filtLF1} and cover the redshifts $z=\{9, 10, 12\}$. If a particular redshift is not depicted it is because there was no flux in the bandpass. For each of these plots we indicate the JWST magnitude cutoff for the deep campaign, 31.4 mag, at redshifts $z=9$ and 12 using dark and light grey regions, respectively.

The HST F125W filter, due to Lyman forest absorption, was unable to detect any of our galaxies at $z > 10$. In fact, at $z=12$, F125W samples across the Lyman limit. However, at $z=9$ our data agrees with the predicted Schechter faint end slope, while the $z=10$ prediction is about 1 dex below the extrapolated Schechter function. However, even at $z=10$ this filter samples across the Lyman-$\alpha$ line and the flux has been attenuated by the intergalactic medium (IGM). 

Examining the data for F160W, we see our simulated galaxies are somewhat bright at $z=9$, but within $\approx 1\sigma$ of the model, while our $z=12$ data are lower than predictions. This is also due to Lyman forest absorption in this bandpass. However, this level of agreement with Schechter functions based on Hubble deep-field surveys at $z=9$ and 10 is evidence that our simulation is producing reasonable results out to these redshifts. 

The situation is similar for the JWST bandpass filter at 1.5 $\mu$m (F150W). Our data for galaxies at $z=9$ and 10 follow the extrapolated Schechter function but once again, at $z=12$, this wide-band filter samples mostly blue-ward of the Lyman-$\alpha$ line in the rest-frame. Hence we are seeing the attenuation of UV photons by the IGM as we go from $z=10$ to 12.

The remaining plots for the JWST filters redder than 1.5 $\micron$ match the predicted Schechter functions well for $z=9$ and 10. The predictions for $z=12$ are slightly lower than would be predicted by extrapolated Schechter functions, however we are in a cosmic era with unobserved and unmeasured galaxy counts as well as a region in which the simulation may be underestimating the numbers of these early structures due to limited resolution. 

Considering our magnitude limit of $m_{\rm AB} = 31.4$ mag, galaxies at $z > 12$ have to be brighter than $M_{\rm AB} \approx -16.4$ to be detected by JWST. We note that none of our simulated galaxies at $z=15$ are detectable given our assumption of a limiting magnitude $m_{\rm AB} = 31.4$ mag. Of course, our relatively small simulation volume did not generate any of the more rare, yet bright, galaxies at these high redshifts. However, filters 2 $\micron$ and redder indicate detections for our galaxies out to $z=12$ -- if just barely. 

\section{Conclusions}
We have used a large-scale cosmological simulation to study high-redshift galaxies and the prospect of finding Pop III-bright galaxies. While several of our contemporaries have done similar work \citep{2017arXiv170202146C, 2017arXiv170102749B, 2016MNRAS.462..235L, 2016ApJ...823..140X, 2015ApJ...807L..12O}, our approach is novel in that our models include the enhancement to Pop III star formation we expect due to the timescale required to turbulently mix pollutants at subgrid scales. We find that our Pop III SFRD is approximately twice what we would have expected without modeling the subgrid pristine fraction of gas. As a check, we have analyzed more than 20,000 galaxies in our simulation volume of 4828 comoving Mpc$^{3}$ producing UV LFs and statistics on the fraction of Pop III-bright galaxies across a range of redshifts. We have also generated LFs for several HST and JWST filters. 

The current observational constraints on $z\ge 8$ LFs are uncertain at best \citep{2016PASA...33...37F, 2015MNRAS.450.3032M, 2015ApJ...803...34B, 2015ApJ...808..104O}.  Determining the faint end slope, $\alpha$, is the challenge here since observations of galaxies dimmer than $M^{*}$ are likely to dominate galaxy number densities at high-redshift and, more importantly, to be the home of Pop III galaxies. We find that linear extrapolations of the faint end slope to $z>8$, as captured in Table \ref{tab:schecParams}, appear reasonable to $z=12$.  While the Schechter function indicates an ever-increasing number of faint galaxies, we know that the actual LF must flatten and turn-over at some point. Even though the simulation's resolution limits our ability to estimate this turn-over magnitude, we have determined that galaxies down to $M_{\textrm UV} = -14$ reasonably follow the extrapolated $\alpha$. Additionally, our simulation demonstrates that $M^{*}_{\textrm UV}$, the absolute magnitude where galaxy counts begin to rapidly decay, is brighter than $M_{\textrm UV} = -16$ out to $z=12$, again in agreement with linear extrapolations of current observations.

The mass-metallicity relation for our simulated galaxies follows the expected trend of increasing metallicity with increasing mass.  When considering galaxies composed purely of Pop III stars, we note that they are very rare and typically have $M_{\rm G} < 10^{9}\, M_{\odot}$. The peak of Pop III galaxy formation occurs immediately before reionization at $z=9$ and 10 where $\approx$17\% and 25\%, respectively, of simulated Pop III galaxies with $Z < Z_{\rm crit}$, have masses $M_{\rm G} > 10^{9}\, M_{\odot}$. 

Turning to Pop III-bright galaxies with at least 75\% of their flux coming from Pop III stars, roughly 17\% of all galaxies brighter than $m_{\rm UV} = 31.4$  mag (\textit{observable} galaxies) are Pop III-bright at $z=9$, immediately before reionization. Less than 3\% of observable galaxies are Pop III-bright between $7\leq z \leq8$, after reionization. Moving to $z=10$, the Pop III-bright fraction falls to 5\% -- a smaller fraction of the set of more luminous observable galaxies. Finally, at $z > 10$, we do not find any galaxies that
are Pop IIIÐbright with $m_{\rm UV} \le 31.4$ mag within our volume. However, we find at least 15\% of galaxies at $z=12$ and 13\% at $z=14$ are Pop IIIÐbright when considering $m_{\rm UV} = 33$ mag, an intrisic magnitude limit within reach of the JWST using lensing. Thus we predict that the best redshift to search for luminous Pop III-bright galaxies is just before reionization, while lensing surveys for fainter galaxies should push to the highest redshifts possible.

Although our simulation's enhanced Pop III SFRD has only minor implications for the LFs, it does play a significant role in the fraction of Pop III flux coming from our observable ($m_{\rm UV} \le 31.4$ mag) high-redshift galaxies. In fact, when we consider the evolution of the subgrid pristine fraction, the fraction of observable Pop III-bright galaxies in the range $9 \le z \le 10$ is $\approx$ 2 times higher than in the classical Pop III case, in which Pop III stars are only generated in cells with gas $\overline{Z} < Z_{\rm crit}$.  This emphasizes the importance of modeling Pop III star formation accurately, since it has a large effect on the types of galaxies we expect to detect at high redshift.  

While our subgrid model greatly improves the code's ability to reliably produce results for a given physical model, we note that other simulations of high-redshift galaxies may make different assumptions about the relevant physics that lead to different conclusions about the observability of PopIII galaxies at $z > 10$ (e.g. \cite{2015MNRAS.446..521S,2014MNRAS.444.3288J,2012ApJ...745...50W}).  For example, a recent simulation by \cite{2015MNRAS.452.1152J} followed the assembly of a single $10^{8}\, M_\odot$ halo in a zoom simulation with a high resolution 300 kpc$^3$ comoving box.  They found that Pop III star formation was subdominant by $z \approx 13$ in this environment and negligible by $z=10$. While some differences from our results are likely due to parameter choices and the type of region being simulated, they also noted that  radiative transfer and related heating played a crucial role in determining their results. While our work handles cooling by molecular hydrogen along with a simple model for H$_2$ photodissociation, we have not yet included radiative feedback, leaving this to future work.  Thus the debate is ongoing as to the relative importance of different aspects of the physics as well as the values for loosely constrained parameters. 

However, our data predict good news for the JWST. Although we have not considered the effects of  attenuation due to dust absorption, our simulation exhibits galaxy counts per magnitude that meet or exceed current, observationally-based predictions for filters redder than $\approx 1.25 \mu$m through $z=10$. 

While the simulation parameters used in this work are only a starting point for modeling the first galaxies, future work will address the sensitivity of these results across a range of values. These results will help guide future searches for Pop III galaxies.

\acknowledgments
We would like to thank Gabriela Huckabee for performing some of the research needed to reduce certain aspects of our simulation data and Mark Richardson for help with \textsc{ramses}. We would also like to thank Rogier Windhorst for helpful discussions concerning the metallicity-mass relation and the high-redshift luminosity function. This work was supported in part by the National Science Foundation under Grants AST-1715876 \& PHY-1430152 (JINA Center for the Evolution of the Elements), and NASA theory grant NNX15AK82G. The simulations and much of the analysis for this work was carried out on the  NASA Pleiades supercomputer. We would also like to thank the NASA High-End Computing Capability support team.

\software{\textsc{ramses} \citep{2002A&A...385..337T}, AdaptaHop \citep{2004MNRAS.352..376A}, MUSIC \citep{2013ascl.soft11011H}, pynbody \citep{2013ascl.soft05002P}, yt \citep{2011ApJS..192....9T}}

\clearpage
\pagebreak

\section{Appendix}
\begin{figure}[h]
\begin{center}
\includegraphics[width=1\columnwidth]{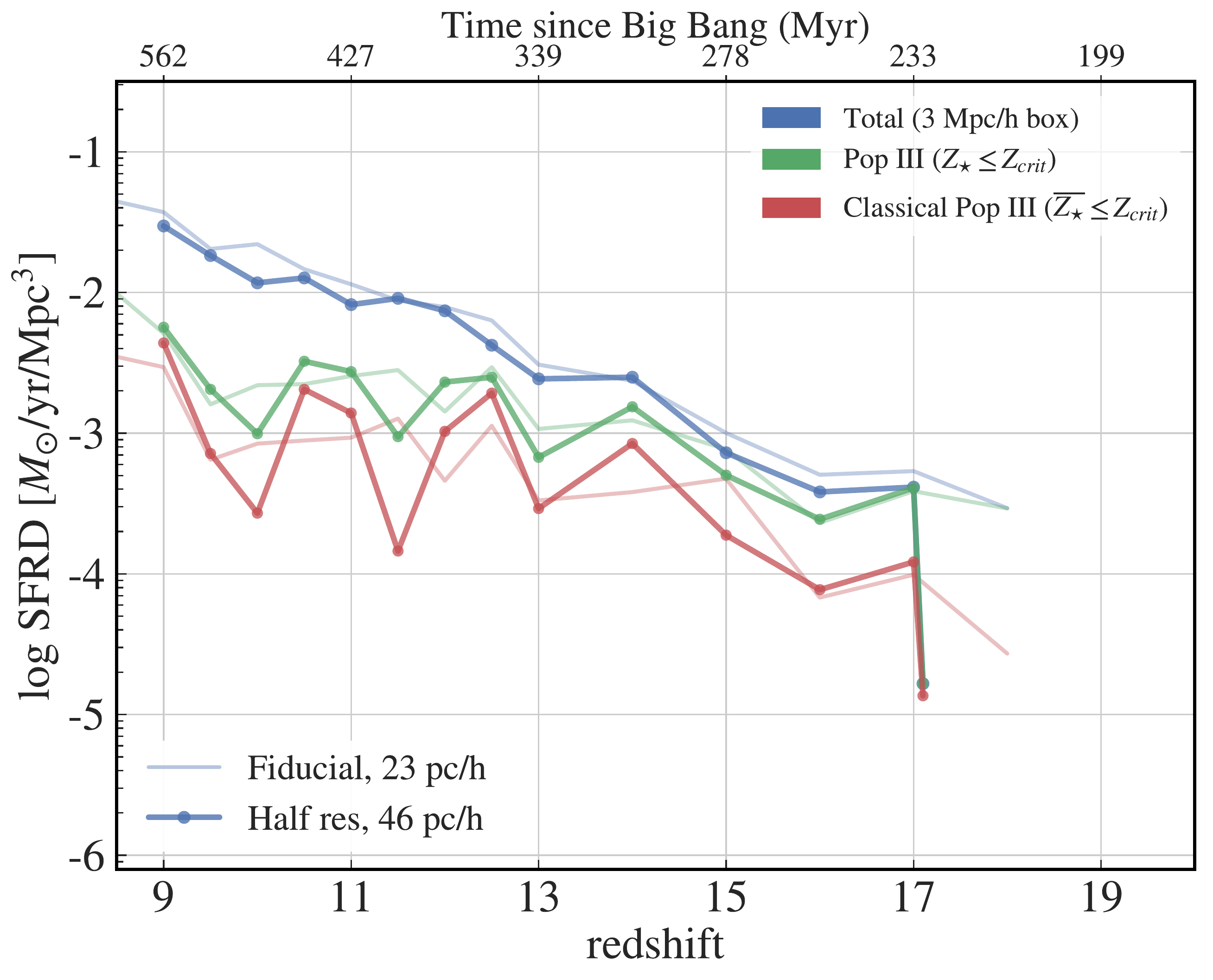}
\caption{The SFRD for the fiducial run in \cite{2017ApJ...834...23S} and a run performed at half of that resolution. While there are inevitable differences between simulations due to the different resolutions, the subgrid model successfully recovers the Pop III rate shortly after the start of star formation at $z\approx18$. This demonstrates that modeling the subgrid fraction of pristine gas effectively improves the resolution of Pop III star formation for the simulation.}
\label{fig:comp}
\end{center}
\end{figure}

In this section we compare the SFRDs from two 3 h$^{-1}$ Mpc$^3$ simulations at different resolutions to demonstrate that the subgrid mixing model described in \cite{2017ApJ...834...23S} -- and used herein -- consistently models the formation of Pop III stars in gas with $Z < Z_{\rm crit}$. The simulation from that work has a average physical spatial resolution of 23 pc h$^{-1}$ resulting in the fiducial SFRD depicted in Figure \ref{fig:comp}. As expected, reducing the average physical resolution to 46 pc h$^{-1}$ results in a delayed and lower SFRD early, since small-scale overdensities are `smoothed over' at lower resolution. However, both the overall SFRD and the Pop III SFRD recover and reach the fiducial level of star formation by $z=16$, demonstrating the subgrid model produces results that converge for Pop III star formation when using different resolutions. 
%

\clearpage 
\pagebreak

\end{document}